\documentclass[aps,prd,superscriptaddress,amsmath,twocolumn,10pt]{revtex4}

\usepackage{color,hangcaption,hhline,psfrag,rotating,amssymb}
\usepackage[hang,nooneline]{subfigure}
\usepackage{dcolumn}
\usepackage{verbatim}
\usepackage{bm}

\begin{document}
\title{An estimate of the hadronic vacuum polarization disconnected contribution to the anomalous magnetic moment of the muon from lattice QCD}

\author{Bipasha Chakraborty}
\affiliation{SUPA, School of Physics and Astronomy, University of Glasgow, Glasgow, G12 8QQ, UK}
\author{C.~T.~H.~Davies}
\email[]{christine.davies@glasgow.ac.uk}
\affiliation{SUPA, School of Physics and Astronomy, University of Glasgow, Glasgow, G12 8QQ, UK}
\author{J.~Koponen}
\affiliation{SUPA, School of Physics and Astronomy, University of Glasgow, Glasgow, G12 8QQ, UK}
\author{G.~P.~Lepage}
\affiliation{Laboratory for Elementary-Particle Physics, Cornell University, Ithaca, New York 14853, USA}
\author{M.~J.~ Peardon}
\affiliation{School of Mathematics, Trinity College, Dublin 2, Ireland}
\author{S.~M.~Ryan}
\affiliation{School of Mathematics, Trinity College, Dublin 2, Ireland}

\date{\today}

\begin{abstract}
The quark-line disconnected diagram is a potentially
important ingredient in lattice QCD calculations of the hadronic vacuum polarization 
contribution to the anomalous magnetic moment of the muon.
 It is also a notoriously difficult 
one to evaluate. 
Here, for the first time, we give an estimate of this contribution based on lattice QCD 
results that have a statistically significant signal, albeit at one value of 
the lattice spacing and an unphysically 
heavy value of the $u/d$ quark mass. 
We use HPQCD's method of determining the anomalous magnetic moment by 
reconstructing the Adler function from time-moments of the current-current 
correlator at zero spatial momentum. 
Our results lead to a total (including $u$, $d$ and 
$s$ quarks) quark-line disconnected contribution to $a_{\mu}$ of 
$-0.15\%$ of the $u/d$ hadronic vacuum polarization contribution with an 
uncertainty which is 1\% of that contribution. 
\end{abstract}


\maketitle

\section{Introduction} 
\label{sec:intro}

\begin{figure}
\centering
\includegraphics[width=0.25\textwidth]{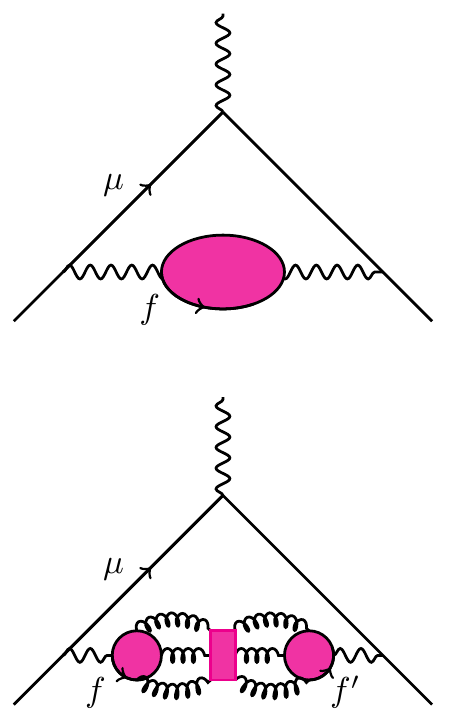}
\caption{
The hadronic vacuum polarization contribution to the 
muon anomalous magnetic moment is represented as a shaded 
blob inserted into the photon propagator (represented 
by a wavy line) that corrects the point-like 
photon-muon coupling at the top of each diagram. 
The top diagram is the connected contribution and 
the lower diagram the quark-line disconnected (but connected 
by gluons denoted by curly lines) contribution
that is discussed here. The shaded box in the lower diagram 
indicates strong interaction effects that could occur between 
the two quark loops.   
}
\label{fig:hvp}
\end{figure}

The high accuracy with which the magnetic moment of the muon 
can be determined in experiment makes it a very useful quantity 
in the search for new physics beyond the Standard Model. 
Its anomaly, defined as the fractional difference 
of its gyromagnetic ratio from the naive value of 2 ($a_{\mu}=(g-2)/2$) is known 
to 0.5 ppm~\cite{Bennett:2006fi}. 
The anomaly arises 
from muon interactions with a cloud of virtual 
particles and can therefore probe the existence of particles 
that have not been seen directly. The theoretical calculation of $a_{\mu}$ in the Standard 
Model shows a discrepancy with the experimental result of 
about $25(8) \times 10^{-10}$~\cite{Aoyama:2012wk, Hagiwara:2011af, Davier} 
which could be an exciting indication of new physics.  
Improvements by a factor of 4 in the experimental uncertainty 
are expected and improvements in the theoretical determination 
would make the discrepancy (if it remains) really compelling~\cite{snowmass}. 

The current theoretical uncertainty is dominated by that from 
the lowest order ($\alpha^2_{\mathrm{QED}}$) hadronic vacuum polarization (HVP) 
contribution, in which the virtual particles are strongly 
interacting, depicted in Fig.~\ref{fig:hvp}.   
This contribution, which we denote $a_{\mu,\mathrm{HVP}}$, is 
currently determined most accurately from 
experimental results on $e^+e^- \rightarrow$ hadrons 
or from $\tau$ decay to be of order $700 \times 10^{-10}$ 
with a 1\% uncertainty or better~\cite{Hagiwara:2011af, Davier, Benayoun:2015gxa}. 
This method for determining $a_{\mu,\mathrm{HVP}}$ does 
not distinguish the two diagrams of Fig.~\ref{fig:hvp} because
it uses experimental cross-section information, effectively 
including all possibilities for final states that would 
be seen if the two diagrams were cut in half. 

$a_{\mu,\mathrm{HVP}}$ can also be determined from lattice QCD calculations
using a determination of the vacuum polarization function at Euclidean-$q^2$ 
values~\cite{Blum:2002ii}. It is important that this is done to 
at least a comparable level of uncertainty to that obtained from 
the experimental results to provide a first-principles constraint of the 
values above. It is hoped that such calculations will, in time, allow the 
theoretical uncertainty to be reduced further. 

Huge progress has been made in lattice QCD 
calculations in the last few years so that accuracies of a few percent in 
$a_{\mu,\mathrm{HVP}}$ are now achievable~\cite{Burger:2013jya}. 
Indeed, a 1\% determination of 
the $s$-quark contribution has been demonstrated~\cite{Chakraborty:2014mwa}. 
These calculations currently include 
only the quark-line connected contribution to the HVP, from 
the top diagram of Fig.~\ref{fig:hvp}. The quark-line disconnected 
contribution, from the lower diagram of Fig.~\ref{fig:hvp},  
vanishes in the SU(3) limit but could still contribute 
several percent to $a_{\mu,\mathrm{HVP}}$ at 
physical $u$, $d$, and $s$ quark masses. 
It cannot therefore be left undetermined if 1\% 
accuracy in $a_{\mu,\mathrm{HVP}}$ is to 
be achieved from lattice QCD calculations. 

Quark-line disconnected diagrams are notoriously difficult 
to evaluate in lattice QCD because of poor signal-to-noise 
properties. Calculations of the disconnected contribution to  
$a_{\mu,\mathrm{HVP}}$ have concentrated on stochastic determinations using 
various noise-reduction methods and several calculations 
are underway, see, for example Ref.~\cite{Francis:2014hoa}. 

Here we use instead lattice QCD results from the Hadron Spectrum 
Collaboration's programme of calculations using 
distillation~\cite{Peardon:2009gh, Dudek:2013yja} in the light quark sector. 
These have enabled a clear signal to be obtained for quark-line disconnected 
correlators. 
Instead of using stochastic methods they rely on computing the correlator 
directly using sources made from a basis of vectors spanning the space of 
the smoothest quark fields.  We combine this approach with HPQCD's method of 
determining $a_{\mu,\mathrm{HVP}}$ by reconstructing the polarization function 
from its $q^2$-derivatives obtained 
from time-moments of correlators at zero spatial momentum~\cite{Chakraborty:2014mwa}. 
HPQCD's approach enables existing meson correlators 
generated for determination of the spectrum, such as those of the 
Hadron Spectrum Collaboration, to be re-used 
for the determination of $a_{\mu,\mathrm{HVP}}$. 
Since the quark-line disconnected vector current correlator 
has been determined using distillation we normalise with the 
$\rho$ meson correlation function using the same method. 

In Section~\ref{sec:method} we give details of the 
method for determining $a_{\mu,\mathrm{HVP}}$ from 
lattice QCD correlators. This method leads to a simple 
(over)-estimate of the disconnected contribution using the 
physical properties of the $\rho$ and $\omega$ mesons given 
in Section~\ref{sec:estimate}. 
In Section~\ref{sec:results}, we give the more complete results 
obtained from the Hadron Spectrum correlators. 
In Section~\ref{sec:discussion} we discuss sources 
of systematic uncertainty in the results that lead
finally in Section~\ref{sec:conclusions} to 
a robust estimate of the impact of the 
disconnected contribution to $a_{\mu,\mathrm{HVP}}$ at the physical point. 

\section{Determining $a_{\mu,\mathrm{HVP}}$ from current-current correlators}
\label{sec:method}

The contribution to the muon anomalous magnetic moment 
from the HVP is obtained 
by inserting the quark vacuum polarization into the photon propagator~\cite{Blum:2002ii, Lautrup:1971jf}: 
\begin{equation}
a_{\mu, \mathrm{HVP}}^{(\mathrm{f}\mathrm{f}^{\prime})} = \frac{\alpha}{\pi} \int_0^{\infty} dq^2 f(q^2) (4\pi\alpha Q_{\mathrm{f}}Q_{\mathrm{f}^{\prime}}) \hat{\Pi}_{\mathrm{f}\mathrm{f}^{\prime}}(q^2),
\label{eq:amu}
\end{equation}
where $\mathrm{f}$ and $\mathrm{f}^{\prime}$ refer to the quark flavours at 
the two ends of the 
polarization function. These two flavours need not be the same when we include the quark-line disconnected 
contribution from the lower diagram of Fig.~\ref{fig:hvp}. 
Here $\alpha \equiv \alpha_{\mathrm{QED}}$ and $Q_{\mathrm{f}}$ is the electric charge of quark $\mathrm{f}$ in units 
of $e$. The function $f(q^2)$ is given by  
\begin{equation}
f(q^2) \equiv \frac{m_{\mu}^2 q^2 A^3 (1-q^2A)}{1+m_{\mu}^2q^2A^2} 
\label{eq:f}
\end{equation}
where
\begin{equation}
A \equiv \frac{\sqrt{q^4 + 4 m_{\mu}^2q^2} - q^2}{2m_{\mu}^2q^2}. 
\label{eq:A}
\end{equation}
The behaviour of $f(q^2)$ means that the integral of 
eq.~(\ref{eq:amu}) is dominated by small values of 
$q^2$ ($\approx m^2_{\mu}$) and hence it is the behaviour of $\hat{\Pi}$ at 
values of $q^2$ close to zero that needs to be determined 
in lattice QCD. 

The quark polarization tensor is the Fourier transform 
of the vector current-current correlator. 
For spatial currents at zero spatial momentum 
\begin{equation}
\Pi^{ii}_{\mathrm{f}\mathrm{f}^{\prime}}(q^2) =
q^2{\Pi}_{\mathrm{f}\mathrm{f}^{\prime}}(q^2) = a^4 \sum_t e^{iqt} \sum_{\vec{x}}\langle j^{i}_{\mathrm{f}}(\vec{x},t)j^{i}_{\mathrm{f}^{\prime}}(0) \rangle 
\label{eq:pi}
\end{equation}
with $q$ the Euclidean energy. We need the renormalized 
vacuum polarization function, $\hat\Pi(q^2)\equiv\Pi(q^2)-\Pi(0)$, 
which automatically removes non-zero contributions to $\Pi(0)$ from  non-conserved 
vector currents. 
Time-moments of the correlator give the derivatives 
at $q^2=0$ of $\hat{\Pi}$
(see, for example, \cite{Allison:2008xk,McNeile:2010ji}): 
\begin{eqnarray}
G_{2n,\mathrm{f}\mathrm{f}^{\prime}} &\equiv& a^4 
\sum_t \sum_{\vec{x}} t^{2n} Z_V^2 \langle j^{i}_{\mathrm{f}}(\vec{x},t)j^{i}_{\mathrm{f}^{\prime}}(0) \rangle  \nonumber \\
&=& (-1)^n \left. \frac{\partial^{2n}}{\partial q^{2n}} q^2\hat{\Pi}_{\mathrm{f}\mathrm{f}^{\prime}}(q^2) \right|_{q^2=0} .
\label{eq:G}
\end{eqnarray}
Here we have allowed for a renormalization factor $Z_V$ for the lattice vector 
current. Note that  time-moments remove any contact terms between the two 
currents. $G_{2n}$ is easily calculated from lattice QCD correlators, 
remembering that $t$ is zero at the origin 
and takes positive values in the positive time direction (up to $T/2-1$) and negative 
values in the negative time direction (down to $-T/2+1$). 

Defining 
\begin{equation}
\hat{\Pi}(q^2) = \sum_{j=1}^{\infty} q^{2j} \Pi_j  
\label{eq:pihat}
\end{equation}
then 
\begin{equation}
\Pi_j = (-1)^{j+1} \frac{G_{2j+2}}{(2j+2)!} \, .
\label{eq:derivs}
\end{equation}
To evaluate the contribution to $a_{\mu}$ we will 
replace $\hat{\Pi}(q^2)$ with its $[n,n]$ and $[n,n-1]$ Pad\'{e} approximants
derived from the~$\Pi_j$~\cite{Pade}.
We perform the $q^2$ integral numerically. 

This method was tested for the connected $s$-quark contribution 
($\mathrm{f}=\mathrm{f}^{\prime}=s$ and including only the 
Wick contractions of the top diagram of Fig.~\ref{fig:hvp})
in~\cite{Chakraborty:2014mwa}, showing that an accuracy of 1\% could be readily achieved 
in that case. 
The Highly Improved Staggered Quark (HISQ) formalism~\cite{HISQ_PRD} was used on improved gluon field configurations 
that include the effect of $u$, $d$, $s$ and $c$ HISQ sea quarks at multiple 
values of the lattice spacing, multiple values of the 
$u/d$ quark mass including the physical value, and multiple 
volumes. Calculations for the connected $u/d$ quark 
contribution are currently underway~\cite{BipashaLAT15}. 

Here we focus on the quark-line disconnected 
contribution to $\hat{\Pi}$, but using existing 
correlators calculated by the Hadron Spectrum collaboration~\cite{Dudek:2013yja} to obtain a result. 
Details are given in the Section~\ref{sec:results}. We first give a simple estimate 
for the contribution based on experimental information about light vector 
mesons. 

\section{An estimate of the disconnected HVP contribution}
\label{sec:estimate}

The quark-line disconnected contribution is shown in
the lower diagram of Fig.~\ref{fig:hvp}.
We need only consider the cases $\mathrm{f},\mathrm{f}^{\prime} \in u, d, s$ 
since quark-line disconnected contributions for 
heavy $c$ and $b$ quarks are suppressed by powers 
of the heavy quark mass~\cite{Kuhn:2007vp}. 
Including the electric charge factors then 
makes clear that the total quark-line disconnected 
contribution to the HVP would vanish in 
the SU(3) limit because 
$\sum_{u,d,s} Q_{\mathrm{f}} = 0$~\cite{Blum:2002ii}. 

Away from this limit, but with 
$m_u=m_d=m_{l}$, the result will be suppressed 
by quark mass factors that are, for example, powers of $m_s-m_l$.  
When the $u$ and $d$ currents are combined with their 
electric charge factors, a light quark current, $j^i_l$, with 
charge factor +1/3 results.  
The total quark-line disconnected contribution 
can then be considered as coming from the quark-line disconnected 
correlator in eq.~(\ref{eq:G}) of a current,
$j^i = j^i_s-j^i_l$. The electric charge associated with 
this combination of currents is 1/3 so that a factor 
of 1/9 appears in eq.~(\ref{eq:amu}). 
This demonstrates a further suppression of the quark-line disconnected 
contribution compared to the connected one, since the 
connected result for $j^i_l$ has an effective 
electric charge factor squared of 4/9+1/9=5/9.  

Three quark-line disconnected correlators
are needed to evaluate the quark-line disconnected 
contribution to the HVP. 
We denote these as $D^{ll}$, ${D}^{ss}$ 
and ${D}^{ls}$ (equal to ${D}^{sl}$), borrowing 
notation from Ref.~\cite{Dudek:2013yja}, where 
the superscripts denote the quark flavours at source and sink. 
The total result is obtained from time-moments of the combination:
\begin{equation}
\label{eq:comb}
D = {D}^{ll}+{D}^{ss} - 2{D}^{ls} .
\end{equation}
This vector current must be renormalized, 
as indicated in eq.~(\ref{eq:G}), and this is achieved by taking ratios with 
the connected correlator made of light quarks, $C^{ll}$. Hence we calculate 
ratios of time-moments of the quark-line disconnected correlator to
those of the connected light quark correlator. The contribution 
to $a_{\mu,\mathrm{HVP}}$ is consequently given as a ratio to that of 
the (dominant) connected light quark contribution. 

As we shall see in Section~\ref{sec:results}, the dominant piece of $D$ is 
$D^{ll}$ and, because $D^{ls}$ has the opposite sign in eq.~(\ref{eq:comb}), 
an overestimate of the magnitude of the disconnected contribution 
to the HVP is obtained from $D^{ll}$ alone. $2D^{ll}$ is the 
difference between the isoscalar and isovector vector correlators. 
At time $t$ larger than the inverse of excited vector meson masses 
(in fact the isovector correlator is saturated by the $\rho$ rather quickly)  
\begin{equation}
\label{eq:dll}
2D^{ll} = \frac{f^2_{\omega}m_{\omega}}{2}e^{-m_{\omega}t} - \frac{f^2_{\rho}m_{\rho}}{2}e^{-m_{\rho}t} .
\end{equation}
$f_{\omega}$ and $f_{\rho}$ are the decay constants of the $\omega$ 
and $\rho$ mesons defined by 
$\langle 0 | j^i | V^k \rangle = f_Vm_V\delta^{ik}$.  
From a simple exponential form it is straightforward to calculate the 
time-moments, converting the sum in eq.~(\ref{eq:G}) to an 
integral. Assuming this ground-state dominance, the ratio of the coefficient of $q^{2j}$ in the 
disconnected and connected $\hat{\Pi}(q^2)$ functions is given by: 
\begin{equation}
\label{eq:pijd}
\frac{\Pi_{j,D}}{\Pi_{j,C}} = \frac{1}{2}\left[\frac{m_{\rho}^{2j+2}f_{\omega}^2}{m_{\omega}^{2j+2}f_{\rho}^2}-1\right].
\end{equation}
If we now include the relative electric charge factors and effects from excited states in the 
correlation functions we have 
\begin{equation}
\label{eq:pijdextra}
\frac{(Q^2\Pi_{j})_D}{(Q^2\Pi_{j})_C} = \frac{1}{10}\left[\frac{m_{\rho}^{2j+2}f_{\omega}^2}{m_{\omega}^{2j+2}f_{\rho}^2}\frac{(1+\delta_{\omega})}{(1+\delta_{\rho})}-1\right] .
\end{equation}
$\delta_{\rho}$ and $\delta_{\omega}$ include terms such as 
$(f_{\rho^{\prime}}/f_{\rho})^2(m_{\rho}/m_{\rho^{\prime}})^{2j+2}$. 
Since the radial excitations of the $\omega$ and $\rho$ are relatively heavy, 
with masses approximately double the ground-state mass~\cite{pdg} and 
we also expect their decay constants to be smaller than those of 
the ground-state, $\delta_{\rho}$ 
and $\delta_{\omega}$ are of order a few percent.  
Within the accuracy of this estimate, they can be ignored.
Since excited $\omega$ masses are in fact typically smaller than excited $\rho$ masses 
we might expect $\delta_{\omega} < \delta_{\rho}$ and so neglecting
these corrections is also consistent with overestimating the 
size of the $D^{ll}$ contribution. 

Dropping the $\delta_{\omega}$ and 
$\delta_{\rho}$ terms in eq.~(\ref{eq:pijdextra}), we can 
evaluate this ratio using information from 
experiment. The difficulty is in determining the 
decay constants from experimental 
information on, for example, the leptonic decay rate. 
Because of the large width 
of the $\rho$, taking the standard approach of setting $q^2$ 
of the photon in this decay 
to the $\rho$ mass is not necessarily 
correct~\cite{pdg, Jegerlehner:2011ti}. Instead one really needs an effective theory that includes $\rho$, $\gamma$ 
(and $\pi\pi$, to be discussed below), as is used in the experimental analysis. 
A sign of this problem is that the standard formula for the leptonic 
decay of the neutral $\rho$ to $e^+e^-$ would yield a decay constant of 
217 MeV using the experimental leptonic width, in contrast 
to the value obtained for the electrically charged $\rho$ from the width of $\tau$ decay 
to $\rho \nu_{\tau}$ which is 209 MeV. There is similar uncertainty for the 
$\omega$ coming not from its width but from mixing with the $\rho$ and/or 
$\phi$. A naive application of the standard formula for the $\omega$ leptonic 
width yields a decay constant of 195 MeV. 

To allow for these uncertainties we evaluate eq.~(\ref{eq:pijdextra}) 
with $f_{\rho}=0.21(1)$ GeV and 
$f_{\omega}=0.20(1)$ GeV. With $m_{\rho}=0.775$ GeV and $m_{\omega} = 0.783$ 
GeV then 
\begin{equation}
\label{eq:pijdres}
\frac{(Q^2\Pi_{j})_D}{(Q^2\Pi_{j})_C} = \left\{\begin{array}{cc}
-0.013(13) & \quad j=1, \\
-0.015(12) & \quad j=2,\\
-0.017(12) & \quad j=3. 
  \end{array}\right. 
\end{equation}
The dominant contribution to the integral of eq.~(\ref{eq:amu}) 
comes from the lowest moment, $j=1$. Since there is little variation in 
the size of the relative contribution with moment number we can 
take 0 to $-2\%$ as our estimate of the contribution of 
$D^{ll}$ to $a_{\mu,\mathrm{HVP}}$ compared to that of $C^{ll}$. 

The non-resonant contributions from multi-$\pi$ meson 
states are not included in this estimate. 
The most important of these is the $\pi\pi$ contribution to the isovector 
channel from direct coupling to the vector current. A simple 
scalar QED calculation of this contribution to 
$a_{\mu,\mathrm{HVP}}$ gives $70\times 10^{-10}$ at 
the physical value of $m_{\pi}$, which is approximately 10\% of the 
total HVP contribution. 
Leading-order chiral perturbation theory (i.e. 
including only $\pi\pi$ terms) gives the ratio of the disconnected 
to connected contributions to the HVP as $-1/10$~\cite{DellaMorte:2010aq}. 
This result is in fact immediately evident from eq.~(\ref{eq:pijdextra}) 
since there is no 
$\pi\pi$ contribution to the isoscalar channel. If the `$\omega$' pieces 
of eq.~(\ref{eq:pijdextra}) are set to zero the result is $-1/10$ for 
each $\Pi_j$ and therefore also for the total integral. 
This is not a particularly 
useful estimate, however, because it only applies to a relatively small 
part of the HVP and {\it not} the total light quark contribution. 
Here we can use it to estimate the disconnected piece of non-resonant 
$\pi\pi$ at $-10\%\times 10\% =-1\%$ of the connected HVP contribution 
to $a_{\mu,\mathrm{HVP}}$. 

A more complete effective theory would be needed to combine the 
resonant and non-resonant contributions above. 
However, bearing in mind that including $s$ quarks 
will reduce the disconnected contribution from eq.~(\ref{eq:comb}), 
a reasonable estimate of the total disconnected contribution 
to $a_{\mu,\mathrm{HVP}}$ is 0 to $-2\%$ of the connected contribution. 
We will see in 
section~\ref{sec:results} that a complete determination from quark-line 
disconnected lattice QCD 
correlators, albeit at an unphysically heavy light quark mass, 
gives a much smaller magnitude than this relative 
to the connected contribution, consistent with the picture that 
our estimate is conservative. 

\section{Lattice Results}
\label{sec:results}

\begin{figure}[t]
\centering
\includegraphics[width=0.45\textwidth]{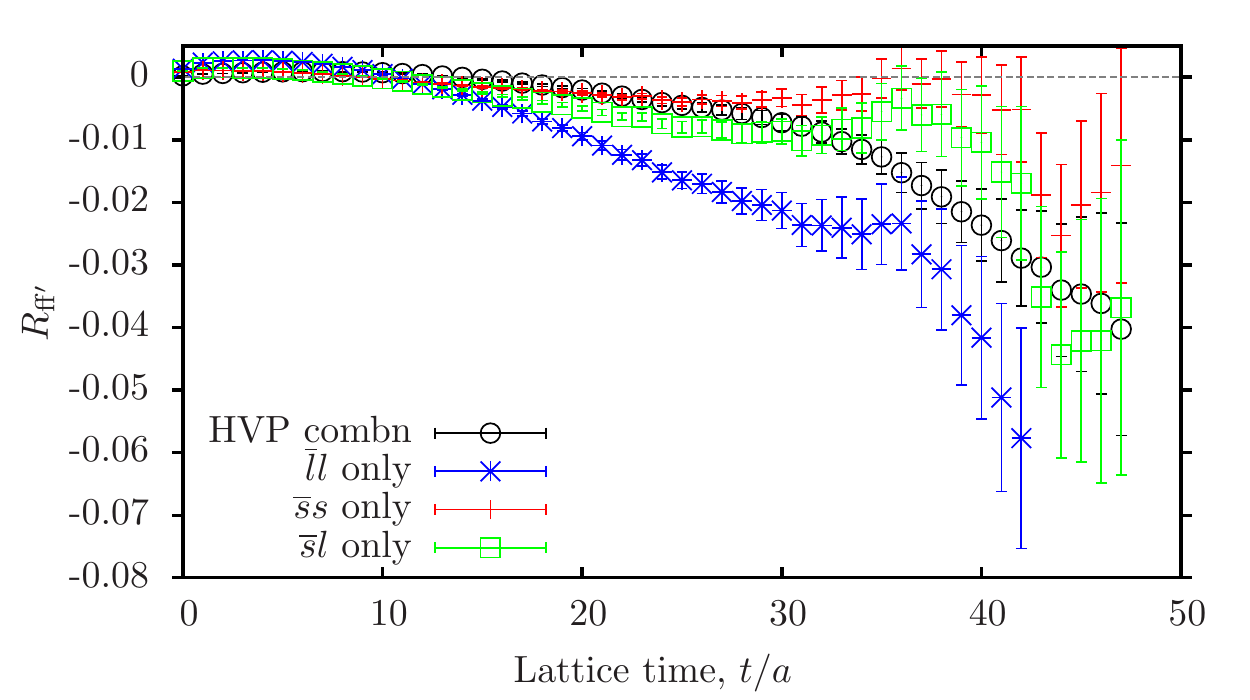}
\caption{ Ratios of disconnected correlators, $D^{\mathrm{ff^{\prime}}}$, 
to the connected correlator $C^{ll}$, as a function of time 
in lattice units. Open black circles show the combination of 
disconnected correlators needed 
for the hadronic vacuum polarization contribution to 
$a_{\mu,\mathrm{HVP}}$, described by eq.~(\ref{eq:comb}).  
}
\label{fig:corr-rats}
\end{figure}

The Hadron Spectrum Collaboration has generated an ensemble of anisotropic gauge field configurations~\cite{Lin:2008pr} with the 
lattice spacing in the temporal direction about 3.5 times smaller than in the spatial directions. The gauge action is tree-level 
Symanzik-improved. The effects of $u, d$ and $s$ quark vacuum fluctuations are included, using a stout-smeared clover quark action. 
The $u$ and $d$ quarks are taken to be degenerate and have a mass approximately one-third 
that of the $s$ quark ($m_{\pi}$ = 391 MeV). The $s$ quark mass 
is tuned to be close to its physical value using the combination of meson masses $2m_K^2-m_{\pi}^2$ fixing the lattice spacing from 
the mass of the $\Omega$ baryon. 
In this study we use the $24^3\times 128$ lattices with mass parameters 
$a_tm_l=-0.0840$ and $a_tm_s=-0.0743$ 
and an inverse temporal lattice spacing of 5.6 GeV. The ensemble consists of 553 configurations. 

The correlation functions employed in this study have a simple local spatial vector 
$\gamma_i$ operator at source 
and sink acting between quark and antiquark fields. The quark bilinear is constructed from 
distilled fields~\cite{Peardon:2009gh} $\tilde{\psi} = \Box \psi$ with 
\begin{equation}
\Box_{xy}(t) = \sum_{k=1}^{N_d} \xi_x^{(k)}(t)\xi_y^{(k)\dag}(t) .
\end{equation}
Here $\xi_k(t)$ is the $k^{\rm th}$ lowest eigenvector of the gauge-covariant three-dimensional Laplace operator on time-slice $t$. In this 
study $N_d=162$. For the disconnected diagrams quark propagation from all time sources is computed.  

The distillation method was developed primarily for hadron spectroscopy applications. Combined with the anisotropic lattices this enabled 
high-resolution and statistically precise determinations of disconnected 
diagrams~\cite{Dudek:2013yja}. 

Using distilled quark fields is not ideal for our
calculation because we wish to determine the time-moments of 
correlation functions constructed from local current operators that 
couple to the photon (as in eq.~(\ref{eq:pi})). We will discuss this 
further below. A smeared correlation 
function, however, has the same exponential behaviour as a local correlation 
function at large times. It simply has a different normalisation for the amplitude. This 
can be fixed if we compare to correlation functions made from the same operator 
whose normalisation we know. Here we can compare the quark-line disconnected 
correlators to the connected correlators to fix the normalisation.    

Figure~\ref{fig:corr-rats} shows the ratio, $R_{\mathrm{f}\mathrm{f}^{\prime}}$, 
of each  
quark-line disconnected correlator $D^{\mathrm{ff^{\prime}}}$ to the connected correlator 
made of light quarks, $C^{ll}$, that uses the same operator at source and sink. 
The figure also includes the ratio for the combination of 
disconnected correlators needed for the 
HVP, as given in eq.~(\ref{eq:comb}). 
Correlators are calculated out to time slice $t=47$, which 
corresponds to 1.6 fm or $7/m_{\rho}$ for these parameters, 
giving ample time for ground-state properties to emerge and 
dominate the connected correlators. 
We see that all of the disconnected contributions become negative 
above a time-slice around 10. Not surprisingly $R_{ll}$ 
has the largest magnitude and $R_{ss}$ the smallest. 
$R_{ss}$ becomes consistent with zero above time-slice 
30, where $R_{sl}$ also becomes small. 
Thus at large times the disconnected contribution to the HVP 
is dominated by the $ll$ component. At shorter times 
there is considerable cancellation between the off-diagonal 
$ls$ piece and the diagonal $ll$ and $ss$ pieces. Directly from 
this figure (and taking into account the factor of 1/5 from electric charge 
factors, see Section~\ref{sec:estimate})  
it is clear that we do not expect the disconnected contribution to 
$a_{\mu,\mathrm{HVP}}$ to amount to more than 1\% of the connected $ll$ contribution.  

In principle to determine the contribution of the disconnected 
correlators to $a_{\mu}$ we simply 
need to determine the time-moments using eq.~(\ref{eq:G}). 
However Figure~\ref{fig:corr-rats} shows that 
the correlators are too noisy at large times for this 
to be a feasible approach. Instead we must fit the correlators 
to their known physical behaviour -- and this requires making combinations 
of connected and disconnected correlators which are physical -- 
and use the fit results at 
large time values. This enables us to make use of the good 
statistical accuracy at short to medium times to fix the long time 
behaviour more precisely. 

We first test this by studying the connected correlators, $C^{ll}$ and 
$C^{ss}$. 
The SU(2) isovector correlator, corresponding to 
flavour combinations $(\overline{u}\gamma_iu-\overline{d}\gamma_id)/\sqrt{2}$, 
$\overline{u}\gamma_id$ and $\overline{d}\gamma_iu$
has no quark-line disconnected contribution in the SU(2) limit. 
The ground-state of the connected light vector correlator $C^{ll}$ is then 
the $\rho$ meson at large times. 
The ground-state of the 
$C^{ss}$ correlator will be a version of the $\phi$ meson in which no mixing 
with other flavourless vector states is allowed. We expect this to be 
very close to the physical $\phi$ meson because $D^{ss}$ is so small. 

We can test the robustness of our correlation function analysis which uses just a single current insertion, 
by comparing to the spectrum analyses of both the Hadron Spectrum and the HPQCD Collaborations. 
A multi-exponential model 
\begin{equation}
\label{eq:fit}
\mathcal{C}_{\mathrm{fit}}(t) = \sum_{i=0}^{\mathrm{n_{exp}}} b_i^2 e^{-E_it} ,
\end{equation}
where $b_i$ and $E_i$ are the amplitudes and masses respectively. 
We use a Bayesian approach~\cite{Lepage:2001ym} to constrain the parameters 
taking a prior of $0.85 \pm 0.6 \,\mathrm{GeV}$ on energy differences 
between the excitations and a width of 0.3 GeV on the ground-state mass. 
The amplitudes are given a prior of $0.1\pm 20$ where the normalisation 
of the correlators is such that the amplitudes of low-lying states 
are around 7--9. Our fit includes the full range of $t$ except for the first 
3 values and stabilises after $\mathrm{n_{exp}}=3$ giving a ground-state mass in 
lattice units of $am_{\rho}=0.1512(4)$ and $am_{\phi}=0.1777(2)$. This is in good 
agreement with the Hadron Spectrum analysis in Ref.~\cite{Dudek:2013yja} which used a large 
number of fermion bilinear operators in a variational basis. The same ensembles were used in a study of 
P-wave $I=1$ $\pi\pi$ scattering which gives a resonance mass of $a_tm_R=0.15085(18)(3)$~\cite{Dudek:2012xn}.
In addition, the value of $m_{\rho}$ at this value of $m_{\pi}$ is close to 
that expected from the HPQCD 
analysis of results at lighter values of $m_{\pi}$~\cite{BipashaLAT15}. 

Using the fits above we can readily determine the $\hat{\Pi}_j$ 
coefficients of eq.~(\ref{eq:derivs}). 
To define a correlation function for any $t$ we combine the calculated correlator 
at short time separations with the model behaviour of eq.~(\ref{eq:fit}). We use 
\begin{equation}
\label{eq:tstar}
C(t) = \left\{ \begin{array}{cc} C_{\mathrm{data}}(t), \quad & t\le t^{\ast}  \\
                               \mathcal{C}_{\mathrm{fit}}(t), \quad  & t > t^{\ast}\end{array}\right. .
\end{equation}
 
From the calculation of the $\hat{\Pi}_j$ we obtain 
the contribution to $a_{\mu,\mathrm{HVP}}$ using eq.~(\ref{eq:amu}), 
with $Q^2_s=1/9$ and $Q^2_l = 5/9$. 
We have tested that the results are insensitive to a number of variations 
of the method. These include: varying $t^{\ast}$ between 20 and 40; 
varying the total time length of the correlator used 
in the calculation of the moments from 95
upwards; varying the number of exponentials used in the fit result
and varying the order of the Pad\'{e} approximant between [1,1] and 
[2,2]. We find the ratio of the $\overline{s}s$ connected contribution to 
$a_{\mu,\mathrm{HVP}}$ to that of the $\overline{l}l$ connected contribution 
to be 0.125. This is in reasonable agreement with a linear 
extrapolation of the HPQCD results 
to the value of $m_{\pi}$ being used here, giving a value
of around 0.15. 

The isoscalar correlator, corresponding to flavour combination 
$(\overline{u}\gamma_iu+\overline{d}\gamma_id)/\sqrt{2}$, has the 
same connected correlator contribution as for the $\rho$ but an 
additional quark-line disconnected contribution of $2{D}^{ll}$. 
The ground-state of this correlator is, to a good approximation, the 
$\omega$ meson. The $\omega$ meson is believed to contain a small 
admixture of $\overline{s}s$ with a mixing angle of a few degrees 
and this is seen in the Hadron Spectrum 
calculations~\cite{Dudek:2013yja}. This mixing occurs via the 
flavour off-diagonal disconnected correlators. We can include 
this effect, as well as establishing the large-time behaviour of 
all the correlators, by simultaneously fitting $ll$, $ss$ and $ls$ combinations of 
correlators with a single set of energy levels as in~\cite{Dudek:2013yja}. 
The $ll$ correlators consist of $C^{ll} + 2D^{ll}$, the 
$ss$ correlators are $C^{ss}+D^{ss}$ and the 
$ls$ correlators are purely quark-line disconnected ($D^{ls}$). 
When $m_u$ and $m_d$ are not equal 
the $\omega$ can also mix with the neutral $\rho$ meson but, since 
we are working with $m_u=m_d$, we neglect this small effect. 

We fit the 3 correlator combinations above
simultaneously, using the fit form given in eq.~(\ref{eq:fit}) 
for the diagonal elements (but with different amplitudes for the $ll$ 
and $ss$ elements) and the form 
\begin{equation}
\label{eq:discfit}
\mathcal{C}^{sl}_{\rm fit}(t) = \sum_{i=0}^{\mathrm{n_{exp}}} d_is_i e^{-E_it} 
\end{equation}
for the off-diagonal element $D^{sl}$, where $d_i$
and $s_i$ do not need to be the same. All combinations share 
the parameters $E_i$. We take very similar priors to our earlier fits. 
However, we change the prior on the energy differences to  
of $700\pm 600 \,\mathrm{MeV}$ to allow for the interleaving of 
excited $\phi$ and $\omega$ levels. 
We also fix a prior on the energy difference between 
the lowest energy (which we expect to correspond 
to the $\omega$) and the second lowest (which we expect to correspond 
to the mass of the $\phi$ meson). This difference is small here because 
the $\omega$ mass is relatively high at these values of $m_l$ (as we saw 
above for the $\rho$), and the $\phi$ mass is slightly lower than its 
physical value. 
We take a prior on the difference of the two energies of 
$170 \pm 100 \,\mathrm{MeV}$. The amplitude prior widths are 
again generally taken to be 20.0. However, we take a smaller prior 
width of 1.0 on $d_i$ and $s_i$ in eq.~(\ref{eq:discfit}), reflecting 
the smaller size of the purely quark-line disconnected pieces. 
We also expect only a weak mixing between $\omega$ and $\phi$ states so 
that the amplitude of the $ss$ combination in the lowest mass state should 
be small and of the $ll$ combination in the second lowest mass state. 
We therefore fix the prior widths of these amplitudes also to be 1.0. 

The fit, using a time range from 3 to 40 is stable from 6 exponentials
upwards with a $\chi^2/\mathrm{dof}=1.1$ (Q=0.2).  
It gives $am_{\omega} =0.1537(10)$ and $am_{\phi}=0.1775(7)$. 
$am_{\omega}$ is somewhat lower than the value (0.1568(4)) obtained by 
the Hadron Spectum collaboration which included mixing between 
light and strange bilinears using more operators and 
the generalised eigenvalue fit method~\cite{Dudek:2012xn}. 
Our $m_{\omega}$ is higher than the corresponding $\rho$ 
mass by 14(6) MeV, compatible
with the physical mass difference of 8 MeV~\cite{pdg}. Note that 
isospin and electromagnetic effects are not included here. 
We obtain the same value of $am_{\phi}$ to that of simply fitting $C^{ss}$, 
so the quark-line disconnected contributions seem to have only a 
small impact, as expected. 

From combinations of the fit results we can determine the large time 
behaviour of the disconnected correlators and therefore their 
time-moments. We again use the correlator data for $t \le t^{\ast}$ 
as in eq.~(\ref{eq:tstar}). For $D^{ll}$ we use correlator data 
up to $t^{\ast}$ and then half the difference of the `$\omega$' 
and `$\rho$' fits. For $D^{ss}$ we similarly take the correlator 
data up to $t^{\ast}$ and then take the difference of the full 
$\overline{s}s$ fit just described, and the result of simply fitting $C^{ss}$ 
above. In fact this gives a very similar result to that of simply including 
the time-moments of the $D^{ss}$ correlator. 
For $D^{sl}$ (=$D^{ls}$) we take the correlator data up to 
$t^{\ast}$ and then the fit result from eq.~(\ref{eq:discfit}).  

Combining this data using eq~(\ref{eq:comb}) with effective charge $Q=1/3$ 
gives a relative contribution to $a_{\mu}^{\mathrm{HVP}}$ of 
\begin{equation}
\label{eq:res}
\frac{a_{\mu,\mathrm{HVP}}^{\mathrm{disc}}}{a_{\mu,\mathrm{HVP}}^{ll,\mathrm{conn}}}=-0.14(5)\% .
\end{equation} 
We have checked that this result is robust (to 50\%) to changing $t^*$; 
the order of the Pad\'{e} approximant and the total time used in 
determining the time-moments. 

The value is made up of -0.36(4)\% from $D^{ll}$, +0.27(3)\% from $D^{ls}$ 
(this has a coefficient of -2 in eq.~(\ref{eq:comb}))
and -0.05(1)\% from $D^{ss}$.  

\section{Discussion}
\label{sec:discussion}

Our result in eq.~(\ref{eq:res}) shows that the disconnected piece of the 
HVP contribution to $a_{\mu}$ is very small. To obtain this result we have 
used vector current-current correlators using distilled quark fields 
and we have worked at rather large 
values of the $u/d$ quark mass at one value of the lattice spacing. 
We discuss each of these issues in turn, bearing in mind that the aim is to 
reduce the uncertainty of the disconnected contributions to the level of 
1\% of the connected contribution. If, as our result indicates, the size 
of the disconnected contribution is already itself of this order, then the relative 
accuracy in the value does not need to be very high. 

\begin{figure}[t]
\centering
\includegraphics[width=0.45\textwidth]{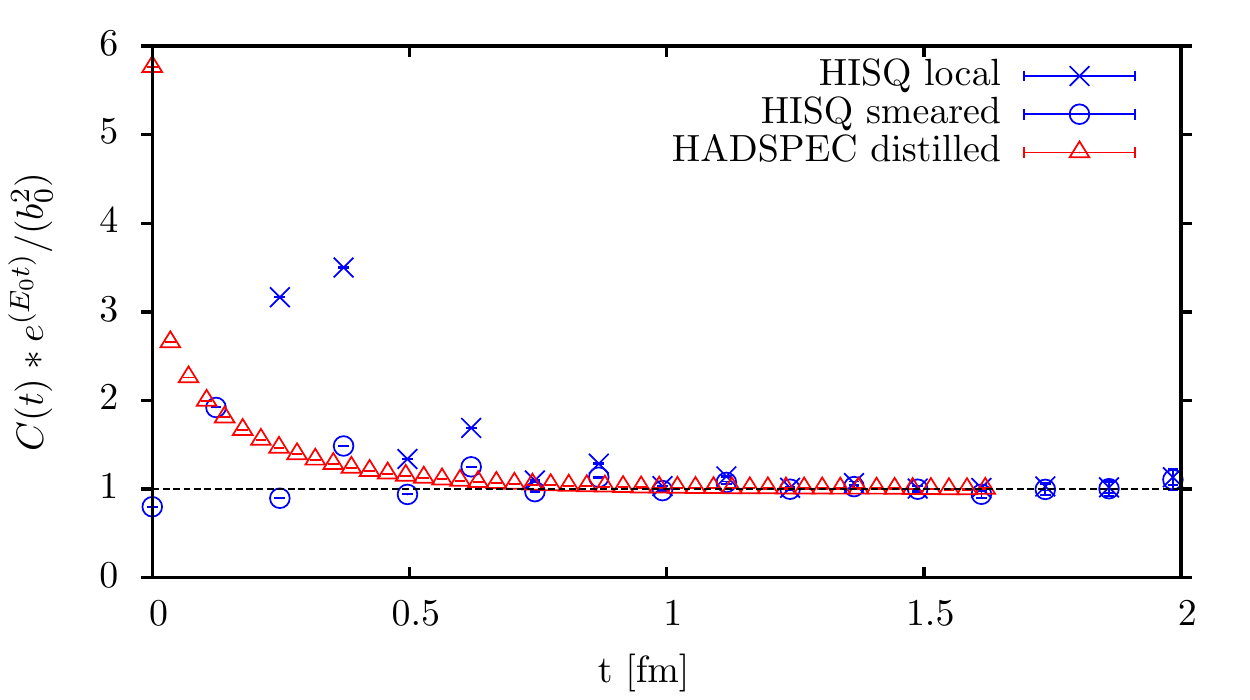}
\caption{ A comparison of local, smeared and distilled vector-vector 
connected correlators normalised to the contribution of the ground-state. 
Results for HISQ $u/d$ quarks compare a local 
operator at source and sink (blue crosses) with 
a smeared operator of Gaussian type and radius 3.75 
at source and sink (blue open circles). 
Red open triangles give the Hadron Spectrum $C^{ll}$ results 
using distilled quark fields from 162 distillation vectors.  
}
\label{fig:compsm}
\end{figure}

To understand the size of systematic error we could be making because of 
the use of distilled quark fields in the vector currents 
we can compare to results from the HPQCD 
collaboration. The HPQCD collaboration has connected $ll$ vector-vector 
correlators for both local and smeared operators using HISQ quarks on a range of gluon 
field ensembles at various values of $m_l$. The reason for using smeared 
operators in this case is to improve the fitted results for the 
ground-state behaviour in the local correlators~\cite{BipashaLAT15}.  
For a comparison between HPQCD and Hadron Spectrum results we want to choose 
an approximately matching spatial lattice spacing, since this controls the 
range of the smearing. The gluon field ensemble used here for the Hadron 
Spectrum results has a spatial lattice spacing of around 0.12 fm, based 
on an anisotropy of 3.444~\cite{Dudek:2012xn}. This corresponds to the 
`coarse' lattice spacing in the MILC gluon field ensembles used by the 
HPQCD collaboration~\cite{Chakraborty:2014mwa}. The MILC coarse ensemble with 
$m_l/m_s = 0.2$ has a value for $m_{\pi}$ of 305 MeV~\cite{fkpi} which is somewhat 
lower than the Hadron Spectrum value used here of 391 MeV, but fairly similar.  
The smearing used by the HPQCD collaboration on these configurations uses a 
stride-2 covariant Laplacian, applying $[1+(r_0^2D^2)/4n]^n$ to a local source. 
$r_0$ is set to 3.75 and $n$ to 30. 
The stride-2 Laplacian is needed to avoid mixing in other staggered tastes of 
vector meson~\cite{HISQ_PRD}. 

Figure~\ref{fig:compsm} compares results for the HPQCD correlators using local 
and smeared operators on these coarse lattices. The quantity plotted is 
the correlator divided by the ground-state contribution so that at large 
times a result of 1 is guaranteed. The $x$-axis gives the time between source 
and sink in fm. We see that smeared correlators are closer to the large 
time behaviour at early times, although this is obscured somewhat by the 
oscillations present in vector correlators made with staggered 
quarks. Since the smearing is designed to increase the 
projection onto the ground-state it is not surprising that smeared correlators 
are saturated earlier by the ground-state. However there is little 
difference between smeared and local correlators 
beyond $t$ = 0.5 fm, equivalent to $2/m_{\rho}$ for this value 
of $m_{\pi}$. We can determine the contribution to $a_{\mu,\mathrm{HVP}}$ 
of the smeared and local correlators, using the method described 
in Section~\ref{sec:results}. If we normalise the two correlators by the 
ratio of their ground-state amplitudes, we find that the smeared correlators 
give a result for $a_{\mu,\mathrm{HVP}}$ that is about 10\% low compared 
to the local result.  

Figure~\ref{fig:compsm} also compares the Hadron Spectrum 
results using distillation.  
Now the density of points in time is much higher, reflecting 
the finer discretisation of the time direction in the Hadron Spectrum 
lattices. The behaviour of the correlators is, however, fairly similar to 
that of the HPQCD smeared correlators (allowing for the oscillations in the 
staggered quark results). This gives some indication of the effective size of the 
Hadron Spectrum smearing. It also implies that we might expect the Hadron 
Spectrum results for contributions to $a_{\mu,\mathrm{HVP}}$ to be 
approximately 10\% low compared to those obtained from local operators.  

To understand to what extent the result might change as $m_l$ and hence 
$m_{\pi}$ is reduced to the physical value, 
we can compare the results 
from the Hadron Spectrum analysis~\cite{Dudek:2013yja} at multiple 
values of $m_{\pi}$ to the picture found in experiment, 
bearing in mind that experiment also has effects from electromagnetism 
and $m_u \ne m_d$ that we are not including. 

The impact of the disconnected correlator contributions to ground-state 
masses is to change 
the mass of the $\omega$ relative to the $\rho$ and to change the 
mixing between the $\phi$ and the $\omega$. The Hadron Spectrum analysis finds, 
even at $m_{\pi} >$ 391 MeV, a picture that is qualitatively and quantitatively 
very similar to experiment, except that $m_{\rho}$ is too heavy.  
The impact of the $\rho$ mass being too heavy is largely removed by 
the fact that we take a ratio of the disconnected contribution to 
that of the connected contribution. 
The $\omega$ is found to be slightly heavier than the $\rho$ and the 
mixing between the $\omega$ and $\phi$ is a few degrees. Excited state masses 
also agree well with experiment. The mass difference 
between $\omega$ and $\rho$ is seen to increase slightly as $m_{\pi}$ falls 
and the mixing angle with the $\phi$ also increases. Large changes are not to 
be expected, however, if the results are to be compatible with experiment 
in the continuum and chiral limits. The masses of the $\eta$ and 
$\eta^{\prime}$, whose correlators have large contributions from 
quark-line disconnected diagrams, also show good (at the 10\% level) 
agreement with experiment. This demonstrates that quark-line disconnected 
contributions are not unduly distorted at heavy values of $m_{\pi}$.

The discussion above relates to meson masses which test the 
correlator time-dependence. We also have to worry about the correlator 
amplitudes which would be tested through the determination of 
decay constants. The most that we can do here is test 
ratios of decay constants because we do not have normalisation 
factors for our
currents. Our fit results for the $\phi$ and $\rho$ yield 
a ratio for the decay constants of 1.03. This is compatible with 
experimental results for the relative leptonic widths, given 
the uncertainties for the $\rho$ discussed in Section~\ref{sec:estimate}. 
Our fit results for the $\rho$ and $\omega$ give decay constants 
that are the same, up to 2\% uncertainties. Again this is compatible 
with experiment. 

The key effect that is $m_{\pi}$-dependent and that 
is being underestimated in these lattice QCD results is 
that of the $\pi\pi$ contribution, both resonant, from $\rho$ decay, and 
non-resonant. From Section~\ref{sec:estimate} we estimated 
that the disconnected $\pi\pi$ contribution to $a_{\mu,\mathrm{HVP}}$
is -1\%. This uncertainty is larger 
than any of the $m_{\pi}$-dependent effects discussed above.   

The fact that only one value of the lattice spacing 
is being used means that we have no direct way of testing 
for discretisation effects. 
The discretisation of QCD used for the Hadron Spectrum results 
has discretisation errors in principle of order $\alpha_s(\Lambda a_s)$ and 
$(\Lambda a_s)^2$. Here $\Lambda$ is a suitable QCD scale that sets the 
size of discretisation effects, say 400 MeV. $a_s$ is the spatial lattice 
spacing, $a_s^{-1}$ = 1.6 GeV. We might therefore expect discretisation 
errors of 5--10\%. 
This is consistent with the comparison to HPQCD results where 
three values of the lattice spacing have been used for connected 
correlator calculations so that a clear continuum limit can be taken 
(and in fact only very small discretisation errors are evident). 
The Hadron Spectrum results for $m_{\rho}$ and $m_{\phi}$ are 
consistent with those from HPQCD at a similar value for $m_{\pi}$ 
within possible 5--10\% discretisation errors. 

We conclude that uncertainties from the effect of distillation 
and the use of relatively heavy $\pi$ mesons at one value of 
the lattice spacing could amount to a total of 50\% of the 
very small value for the ratio of disconnected to connected 
contributions to the HVP found in Section~\ref{sec:results}.  
A larger uncertainty comes from the $\pi\pi$ contributions 
that are badly distorted at heavy $m_{\pi}$ and this will dominate 
our final uncertainty. 

\section{Conclusions}
\label{sec:conclusions}
The ultimate aim of lattice QCD calculations of $a_{\mu,\mathrm{HVP}}$ is 
to improve on results from using, for example, $\sigma(e^+e^- \rightarrow \mathrm{hadrons})$ 
that are able to achieve an uncertainty of below 1\%. 
We are not at that stage yet. The ETM Collaboration are the 
first to include a full calculation from connected correlators 
including $u/d$, $s$ and $c$ quarks~\cite{Burger:2013jya} 
and quote a 4\% uncertainty that includes lattice systematic uncertainties. 
A 1\% uncertainty has now been achieved on the $s$ quark connected 
contribution~\cite{Chakraborty:2014mwa} and 
an improved accuracy on the total $u/d$, $s$ and $c$ quark connected 
correlator calculation is within reach~\cite{BipashaLAT15}. 
However neither of these calculations 
includes the impact of quark-line disconnected correlators. 
Although the total disconnected contribution to $a_{\mu,\mathrm{HVP}}$
is expected to be small, it must either be evaluated or constrained at 
the level of 1\% of the total if it is not to undermine our ability to reach 
the desired accuracy on $a_{\mu,\mathrm{HVP}}$ from lattice QCD calculations. 

Here we have given the first estimates of the quark-line disconnected 
contribution to $a_{\mu,\mathrm{HVP}}$ 
based on disconnected correlators with a clear signal. 
We determine the disconnected contribution as a ratio to the connected 
$u/d$ contribution so that a number of systematic errors cancel or 
are reduced. 
The results show that the disconnected contribution is indeed small, 
at -0.15\% of the connected contribution at the relatively heavy value 
of $m_{\pi}$ (391 MeV) used here. We estimate the uncertainty in   
this contribution as 1\% of the connected contribution coming largely 
from $\pi\pi$ effects that are badly distorted at heavy
$m_{\pi}$. 
The value is consistent with a simple phenomenological bound based on the 
experimental properties of the $\rho$ and $\omega$ mesons which 
gives, again as a ratio to the connected contribution, between 0 and -2\%. 

In future a more accurate calculation of the quark-line disconnected 
contributions will be possible with smaller values of $m_{\pi}$, 
going down to the physical point, and on finer lattices. This should enable 
us eventually to pin down these contributions at the 0.1\% level. 
The result given here however is enough to make clear that 
the quark-line disconnected contribution 
to $a_{\mu,\mathrm{HVP}}$ can safely be assessed to be at the level 
of 1\% of the connected contribution. It will not therefore prevent us, 
for now, in reaching an accuracy on the total $a_{\mu,\mathrm{HVP}}$ 
of around 1\% from lattice QCD. 

{\it Acknowledgements.} 
We are grateful to the Hadron Spectrum Collaboration for the correlators 
used here and to C. Bernard for useful discussions. Chroma~\cite{Edwards:2004sx} and QUDA~\cite{Clark:2009wm,Babich:2010mu} were used on clusters at Jefferson Laboratory under the USQCD Initiative and the LQCD ARRA project. Gauge configurations were generated using resources awarded from the U.S. Department of Energy INCITE program at Oak Ridge National Lab, the NSF Teragrid at the Texas Advanced Computer Center and the Pittsburgh Supercomputer Center as well as at Jefferson Lab. 
Calculations were also done on the Darwin Supercomputer 
as part of STFC's DiRAC facility jointly
funded by STFC, BIS 
and the Universities of Cambridge and Glasgow. 
This work was funded by the Gilmour bequest to the University of 
Glasgow, Science Foundation Ireland (grant 11-RFP.1-PHY-3201), 
the Science and Technology Facilities Council, 
the Royal Society, the Wolfson Foundation
and the National Science Foundation.
\bibliography{g2disc}

\begin{thebibliography}{29}
\bibitem[{Pad()}]{Pade}
\bibinfo{note}{{The $[m,n]$ Pad\'e approximant of a function $f(x)$ is a ratio
  of polynomials in~$x$, of order $m$ in the numerator and $n$ in the
  denominator, whose Taylor expansion is the same as that of $f(x)$ through
  order~$x^{n+m}$.}}
\end{thebibliography}

\end{document}